\documentclass[11pt]{article}

\usepackage{times}
\usepackage{helvet}
\usepackage{courier}
\usepackage{graphicx}
\usepackage{color}
\usepackage{epsfig}
\usepackage{enumerate}
\usepackage{wrapfig}
\usepackage{pgf}

\usepackage{booktabs}

\usepackage{amsmath}
\usepackage{amsfonts}
\usepackage{amsthm}
\usepackage{amssymb}

\newcommand{\pref}{\succ}

\newcommand{\np}{{\mathrm{NP}}}

\newcommand{\calR}{\mathcal{R}}

\newcommand{\pos}{{{{\mathrm{pos}}}}}

\def\ts{\textstyle}

\newcommand{\shortcite}[1]{\cite{#1}}

\title{What Do Multiwinner Voting Rules Do? \\ An Experiment Over the
  Two-Dimensional Euclidean Domain\thanks{This paper has originally
    appeared in proceedings of the AAAI-2017 conference. The current
    version includes additional material in the appendix, a
    correction of a bug  (in the AAAI paper we claimed results for the $\alpha_k$-PAV
    rule, while in fact they were for the HarmonicBorda rule; this is corrected in this
    manuscript), and minor updates to the references.}}
\author{
  \phantom{123456789012345678901}\\
  Edith Elkind\\
  University of Oxford\\
  Oxford, UK \and
  \phantom{12345678901234567}\\
  Piotr Faliszewski\\
  AGH University \\
  Krakow, Poland \and
  \phantom{123456789012}\\
  Jean-Fran{\c c}ois Laslier\\
  Paris School of Economics\\
  Paris, France \and
  Piotr Skowron\\
  University of Warsaw\\
  Warsaw, Poland \and
  Arkadii Slinko \\
  University of Auckland\\
  Auckland, New Zealand \and
  Nimrod Talmon\\
  Ben-Gurion University\\
  Be'er Sheva, Israel }

\begin{document}

\maketitle

\begin{abstract}
  We visualize aggregate outputs of popular multiwinner voting rules---SNTV, 
  STV, Bloc, $k$-Borda, Monroe, Chamberlin--Courant, and HarmonicBorda---for 
  elections generated according to the two-dimensional Euclidean model.
  We consider three applications of multiwinner voting, namely, 
  parliamentary
  elections, portfolio/movie selection, and shortlisting, and 
  use %
  our results
  to understand which of our rules %
  seem to be best suited for each application.
  In particular, we show that STV (one of the few nontrivial rules
  used in real high-stake elections) exhibits excellent performance, 
  whereas the Bloc rule (also often used in practice)
  performs poorly.
\end{abstract}

\setlength{\fboxsep}{0pt}  %
\newcommand{\histogram}[2]{\fbox{\includegraphics[width=1.6cm]{hist_#1200/#2_20_hist.png}}}
\newcommand{\samplerun}[2]{\fbox{\includegraphics[width=1.6cm]{sample_run/#1_#2_20.png}}}

\newcommand{\resultsrule}[2]{
\rotatebox{90}{#1} &
\histogram{gauss}{#2}
\samplerun{gauss}{#2}&
\histogram{circle}{#2}
\samplerun{circle}{#2}&
\histogram{uniform}{#2}
\samplerun{uniform}{#2}&
\histogram{gauss4}{#2}
\samplerun{gauss4}{#2}\\[1mm]
}

\newcommand{\pfparagraph}[1]{\medskip

\noindent\textbf{#1}\quad}

\section{Introduction}

The goal of this paper is to develop a better understanding of a
number of well-known multiwinner voting rules, by analyzing their
behavior in elections where voters' preferences are generated according
to a two-dimensional spatial model.  By focusing on this preference
domain, we can visualize the election results and check if they agree
with the intuition and motivation behind these rules.  Our study can
be seen as an experimental counterpart of the work of of Elkind et
al.~\shortcite{elk-fal-sko-sli:j:multiwinner-properties}, who 
analyze multiwinner rules axiomatically. %

In a multiwinner election, the goal is to select
a size-$k$ committee (i.e., a set of $k$
candidates, where $k\in{\mathbb N}$  is part of the input) based on the voters'
preferences. Usually, voters can express their preferences by listing the candidates
from best to worst or by indicating which candidates they approve;
we focus on the former setting, as it %
fits the spatial preference
model better. %

Applications of multiwinner voting range from choosing a parliament
through preparing a portfolio of company's
products~\cite{bou-lu:c:chamberlin-courant,bou-lu:c:value-directed-cc}
or choosing movies to offer to passengers on a long
flight~\cite{elk-fal-sko-sli:j:multiwinner-properties,sko-fal-lan:j:collective}
to shortlisting runners-up for an
award~\cite{bar-coe:j:non-controversial-k-names,elk-fal-sko-sli:j:multiwinner-properties}.
As a consequence, there is also quite a variety of different
multiwinner voting rules. For instance, for parliamentary elections
an important desideratum is proportional representation of the voters, 
and there are voting rules such
as STV or the Monroe rule (we define all the rules considered in this paper 
in the next section) that have been designed with this idea in mind. 
On the other hand, in the context of portfolio or movie selection
we primarily care about the diversity of the selected committee,
and it has been argued that the Chamberlin--Courant rule 
is good for this purpose \cite{bou-lu:c:chamberlin-courant,sko-fal-lan:j:collective}. 
For shortlisting, our primary concern is fairness:
if there are two similar candidates, we want to select both or neither, 
and increasing the target committee size
should not result in any of the selected candidates being dropped;
these requirements are satisfied by $k$-Borda.
Naturally, there are %
other scenarios which require
other %
normative properties.

The examples above indicate that choosing a good multiwinner rule
is not a trivial task. It is therefore natural to ask 
how we can facilitate the decision-making process of a user who is facing this choice.
There are several good answers to this question. First, some rules are
specifically designed for certain tasks. For example, STV and the Monroe rule
have explicit built-in mechanisms ensuring that every
sufficiently large group of like-minded voters is represented. Second,
we can analyze axiomatic properties of the rules. This line of work,
was extensively pursued for single-winner rules;
for the case of multiple winners in was initiated by Felsenthal and Maoz~\shortcite{fel-mao:j:norms} and
Debord~\shortcite{deb:j:k-borda}, with recent contributions including
the work of Elkind et
al.~\shortcite{elk-fal-sko-sli:j:multiwinner-properties} and Aziz et
al.~\shortcite{azi-bri-con-elk-fre-wal:j:justified-representation}.
Finally, one can use empirical analysis to compare different rules
under particular conditions. For example, Diss and
Doghmi~\shortcite{dis-dog:t:condorcet-scoring-rules} 
consider a few multiwinner voting rules and experimentally investigate how frequently they
pick Condorcet committees.\footnote{In a Condorcet committee, every committee
member is preferred to every non-member by a majority of the voters.}
All these approaches are useful, and the choice of a voting rule should
take all of them into account.

Nonetheless, a non-expert user may still feel ill at ease when
deciding which rule to choose for his or her particular application.
In this case, a picture may be worth a thousand words: 
a simple graph that clearly explains differences between rules
can be very informative. The contribution of this paper 
is to propose a novel approach to selecting a suitable mutiwinner rule,
which is based on graphical information. That is, 
we provide images that we expect to be helpful in discussions of
multiwinner voting rules. Naturally, reality is too complicated for a
single picture to constitute a definite argument, but we believe
that, on the one hand, our results provide good illustrations
confirming intuitions regarding various multiwinner rules and, on the
other hand, they highlight some faults of the rules that otherwise
would not be easily visible.

\pfparagraph{Our Methodology.} 
The outcome of an election depends both on the voting
rule and on the set of candidates. In this work, we focus on the former aspect
and ask what multiwinner rules do when choosing from a set of
candidates that is representative of the electorate, i.e., under what one may 
call the \textit{representative candidacy} assumption.  We evaluate a
number of multiwinner voting rules (SNTV, STV, Bloc,
$k$-Borda, %
Chamberlin--Courant, Monroe, and HarmonicBorda) on elections generated
using the two-dimensional Euclidean model of preferences. In this
model each candidate and each voter is represented by a point on a
plane, and voters form their preference orders by ranking the
candidates that are closer to them above the ones that are further away.

\begin{figure}
  \centering
  \begin{tabular}{cp{0.2cm}c}
  STV && $k$-Borda \\
  \fbox{\includegraphics[width=5cm]{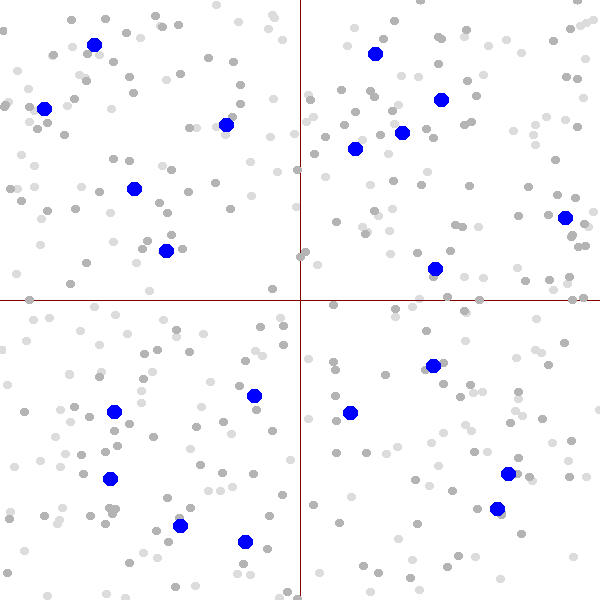}} &&
  \fbox{\includegraphics[width=5cm]{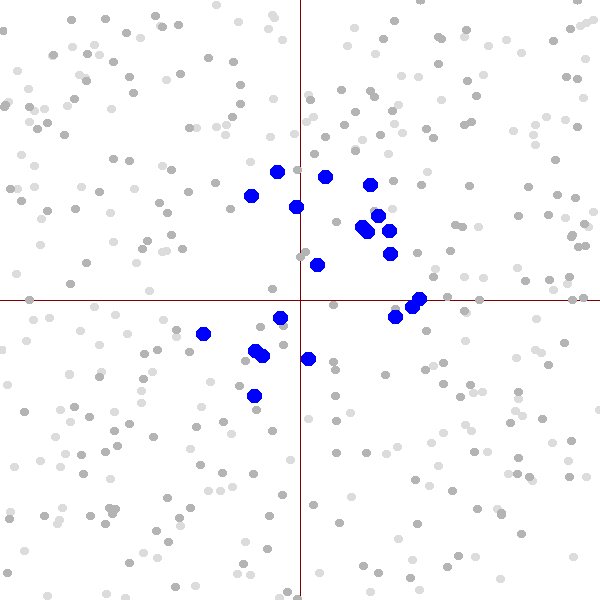}}
  \end{tabular}
  \caption{\label{fig:sample}Results of an election (generated using the 2D
    Euclidean model) according to STV (left) and $k$-Borda (right).
    Voters are depicted as dark gray dots, candidates as light gray
    dots, and the winners as larger blue dots.}
\end{figure}

This model is very appealing and extensively
studied~\cite{enelow1984spatial,enelow1990advances} because of its
natural interpretations: A point representing a candidate or a voter
simply specifies his or her position regarding two given issues.
In the world of politics, these two issues could be, for example, the
preferred levels of taxation and immigration, or the
extent to which the individual believes in personal and economic
freedom. While in some settings more dimensions may be necessary, 
the popularity of the Nolan Chart, which is used to represent 
the spectrum of political opinions, indicates that two dimensions
are often sufficient to provide a good approximation of voters' preferences.

In Figure~\ref{fig:sample} we show a
sample election (the points for candidates and voters are generated
using uniform distribution over a square) and the committees selected
by STV (left) and $k$-Borda (right). It is quite evident that the
committee on the left would form a far more representative parliament
than the one on the right, whereas the one on the right would probably
be a better choice for the set of candidates that are shortlisted for a
position, since they are similar to each other and receive broad support 
among the voters (in particular, no voter ranks them close to the bottom of their list).

Our main contributions are as follows:
\begin{enumerate}
\item For each of our voting rules and four distributions of
  candidates and voters (Gaussian, uniform on a disc, uniform on a
  square, and a mix of four Gaussians), we have generated $10\,000$
  elections and built histograms (Figure~\ref{fig:main}) indicating
  how likely it is that a candidate from a given position will be
  selected.

\item We consider three applications of multiwinner voting, and, for
  each application, we identify the voting rules in our collection
  that are most appropriate for it. We make these recommendations
  based on our histograms and %
  certain statistical properties of the elected committees.  E.g., we
  confirm that STV is an excellent rule for parliamentary elections,
  even superior to the Monroe rule; HarmonicBorda can also be seen as an
  interesting rule that chooses fairly representative committees,
  ignoring candidates with extreme opinions. We also provide evidence
  that Bloc should be treated very carefully since it may not perform
  as well as one might expect (this is particularly important because
  Bloc is among the most popular multiwinner rules).
\end{enumerate}

We present some of our results in the appendix (in particular, this is
the case for the analysis of approximation algorithms for the Monroe
and Chamberlin--Courant rules).

\section{Preliminaries}

For every positive integer $n$, we write $[n]$ to denote the set $\{1, \ldots, n\}$.

\pfparagraph{Elections.} An {\em election} $E = (C,V)$ consists of a set $C =
\{c_1, \ldots, c_m\}$ of {\em candidates} and a list $V = (v_1, \ldots,
v_n)$ of {\em voters}. Each voter $v_i$ has a {\em preference order} $\pref_i$,
i.e., a ranking of the candidates from the most to the least favored
one (according to this voter). For a voter $v$ and a candidate $c$, we
write $\pos_{v}(c)$ to denote the position of $c$ in $v$'s
preference order (where the top-ranked candidate has position~$1$). 
A {\em committee} is a subset of $C$.

A {\em multiwinner voting rule} is a function $\calR$ that, given an election
$E = (C,V)$ and a target committee size $k$ ($1 \leq k \leq |C|$), outputs a
nonempty set of size-$k$ committees; 
these committees are said to {\em tie as election winners}. 
In practice, one has to use some 
tie-breaking mechanism. For our experiments, 
whenever we need to break a tie (possibly at an intermediate stage 
in the execution of the rule), we make a random choice
with a uniform distribution over all possibilities. 

\pfparagraph{(Single-Winner) Scoring Functions.} For an election with $m$ candidates, a {\em scoring function
$\gamma_m$} associates each position $j$, $j\in[m]$, with a {\em score}
$\gamma_m(j)$.  The {\em $\gamma_m$-score that candidate $c$ receives from
voter $v$} is $\gamma_m(\pos_{v}(c))$. The {\em $\gamma_m$-score of
candidate $c$ in election $E$} is the sum of the
$\gamma_m$-scores that $c$ receives from the voters in $E$. We consider the
following two prominent families of scoring functions:
\begin{enumerate}
\item The Borda scoring function, $\beta_m$, is defined as $\beta_m(j) = m-j$.
\item For each $t \in [m]$, the $t$-Approval scoring function, $\alpha_t$, is defined
  as $\alpha_t(j)=1$ if $j \leq t$ and $\alpha_t(j)=0$ otherwise.
  The candidate's $1$-Approval score is known as her Plurality score.

\end{enumerate}

\pfparagraph{Multiwinner Rules.}
We focus on the following multiwinner rules (in the description below
we consider an election $E = (C,V)$, with $m$ candidates and $n$
voters, and committee size $k$):

\begin{description}
\item[SNTV.] The Single Nontransferable Vote rule (SNTV) outputs $k$ candidates
  with the highest Plurality scores.

\item[STV.] The Single Transferable Vote rule (STV) executes a series
  of iterations, until it finds $k$ winners. A single iteration
  operates as follows: If there is at least one candidate with
  Plurality score at least $q = \lfloor \frac{n}{k+1} \rfloor+1$, then
  a candidate with the highest Plurality score is added to the
  committee; then 
  $q$ voters that rank him or her first
  are removed from the election %
  (our randomized tie-breaking plays an important role here), and the selected
  candidate is removed from all voters' preference orders. If there
  is no such candidate, then a candidate with the lowest Plurality
  score is removed from the election (again, ties are broken uniformly at random). 
  The Plurality scores are then recomputed.

\item[Bloc.] Under the Bloc rule we output $k$ candidates with the
  highest $k$-Approval scores (intuitively, each voter
  is asked to name his or her $k$ favorite committee members, and those mentioned 
  most frequently are elected).

\item[$\boldsymbol k$-Borda.] Under the $k$-Borda rule we output %
  $k$ candidates with the highest Borda score.

\item[$\boldsymbol \beta$-CC.] The (classical) Chamberlin--Courant
  rule ($\beta$-CC) is defined as follows \cite{cha-cou:j:cc}. 
  A {\em $k$-CC-assignment function} %
  is a function $\Phi \colon V \rightarrow C$ such that $|\Phi(V)|
  \leq k$ (i.e., $\Phi$ associates each voter with a candidate in a set $W\subseteq C$, $|W|\le k$; 
  for a voter $v$, candidate $\Phi(v)$ is referred to as
  {\em $v$'s representative}).  The {\em $\beta$-CC score} of an assignment $\Phi$ is
  defined as $\beta(\Phi) = \sum_{v \in V} \beta_m(\pos_v(\Phi(v)))$
  (i.e., it is the sum of the Borda scores of voters'
  representatives). $\beta$-CC finds a $k$-CC-assignment $\Phi$ that
  maximizes $\beta(\Phi)$ and outputs the committee $\Phi(V)$ (if it
  happens that $|\Phi(V)| < k$---a situation that occurs, e.g., when all
  the voters have identical preference orders---then $\beta$-CC
  supplements $\Phi(V)$ with $k - |\Phi(V)|$ 
  candidates selected at random).

\item[$\boldsymbol \beta$-Monroe.] The (classical) Monroe rule \cite{mon:j:monroe} is
similar to $\beta$-CC, except that it is restricted to
  $k$-Monroe-assignments. A {\em $k$-Monroe-assignment} is a
  $k$-CC-assignment that satisfies the following constraints: (a)
  $|\Phi(V)| = k$, and (b) for each candidate $c$ such that
  $\Phi^{-1}(c) \neq \emptyset$ (i.e., for each selected
  representative) it holds that $\lfloor \frac{n}{k} \rfloor \leq
  |\Phi^{-1}(c)| \leq \lceil \frac{n}{k} \rceil$.  Intuitively, under
  the Monroe rule each selected candidate represents, roughly, the
  same number of voters.

\item[HarmonicBorda (HB).] The HarmonicBorda rule, introduced by Faliszewski et
  al.~\cite{fal-sko-sli-tal:c:paths} but inspired by the PAV rule
  (see, e.g., the works of
  Kilgour~\shortcite{kil:chapter:approval-multiwinner}, Aziz et
  al.~\cite{azi-bri-con-elk-fre-wal:j:justified-representation}, and
  Lackner and Skowron~\cite{lac-sko:c:thiele-rules} for detailed
  discussions of the PAV rule) operates as follows.
  For a voter $v$ and a committee $W$ such that $v$ ranks the members
  of $W$ on positions $p_1 < \cdots < p_k$, the {\em HarmonicBorda
    score} that $v$ assigns to $W$ is $\mathrm{HB}(W,v) = \sum_{t=1}^k
  \frac{1}{t}\beta_m(p_t)$.  For an election $E = (C,V)$, the {\em HB
    score} of a committee $W$ is defined as HB$(W,V) = \sum_{v \in
    V}\mathrm{HB}(W,v)$.  The rule outputs a committee with the
  highest HB score. 

\end{description}

With our tie-breaking, STV, SNTV, Bloc, and $k$-Borda are computable
in polynomial time using straightforward algorithms. Unfortunately,
the Chamberlin--Courant and Monroe rules are $\np$-hard to compute
(Procaccia et
al.~\shortcite{pro-ros-zoh:j:proportional-representation} show this
for variants of these rules that use $t$-Approval scores $\alpha_t$
instead of $\beta$; for the Borda-based variants defined here, the
results for the Chamberlin--Courant rule and the Monroe rule are due
to Lu and Boutilier~\shortcite{bou-lu:c:chamberlin-courant} and
Betzler et al.~\shortcite{bet-sli-uhl:j:mon-cc}, respectively).  We
compute these rules by solving their integer linear programming (ILP)
formulations (suggested by Lu and
Boutilier~\shortcite{bou-lu:c:chamberlin-courant} for the case of
Chamberlin--Courant, and by Skowron et
al.~\shortcite{sko-fal-sli:j:multiwinner} for the case of Monroe).
$\np$-hardness of the HB rule is folklore and we use a simplified
version of the ILP formulation proposed by Skowron et
al.~\shortcite{sko-fal-lan:j:collective} to compute it (see the
appendix for details).

\pfparagraph{Euclidean Preferences.} Given two points on the plane,
$p_1 = (x_1,y_1)$ and $p_2 = (x_2,y_2)$, we write $d(p_1, p_2)$ to denote
the distance $\sqrt{(x_2-x_1)^2 + (y_2-y_1)^2}$ between them.

In a two-dimensional Euclidean election $E = (C,V)$, each entity $e$
(i.e., either a candidate or a voter) is associated with a point $p(e)
= (x(e),y(e))$. %
Given a pair of candidates $c_i, c_j\in C$, a voter $v\in V$ prefers $c_i$ to $c_j$
if $d(p(v), p(c_i)) < d(p(v),p(c_j))$.
Note that this condition does not constrain voter's preferences over two equidistant
candidates. In our case, since we draw our elections at random, such
situations are unlikely to happen. When they do, we break the tie arbitrarily.

Euclidean preferences are very useful to realistically model political
preferences and, in many cases, to model preferences in shortlisting
tasks. Unfortunately, they are not nearly as useful for modeling
preferences over movies. The reason is that people often do not have a
single most favorite type of a movie, but rather like various genres
for different reasons. Nonetheless, investigating rules meant for the movie
selection application (i.e., for selecting diverse committees) in our framework 
is still important. On the one hand, movie selection is not the only
application where diverse committees are needed, and, on the other
hand, if a rule behaves badly on the Euclidean domain, then it is
unlikely that it would behave well for richer preference models.

\section{Main Results and Analysis}

\pfparagraph{Experimental Setup.} 
We assume that both the candidates and the voters have ideal 
positions in a two-dimensional Euclidean issue space that are
drawn from the same distributions. 
For each voting rule and each distribution,
we generated $10\,000$ elections, each with $m = 200$ candidates and $n = 200$
voters, and for each of them we computed a winning committee of size
$k = 20$.

We consider four distributions of the ideal positions:

\begin{description}
\item[Gaussian.] Ideal points are generated using symmetric Gaussian
  distribution with mean $(0,0)$ and standard deviation~$1$.
\item[Uniform Square.] Ideal points are distributed uniformly on the
  square $[-3,3] \times [-3,3]$.
\item[Uniform Disc.] Ideal points are distributed uniformly on the
  disc with center $(0,0)$ and radius $3$.
\item[4-Gaussian.] Ideal points are generated using four symmetric
  Gaussian distributions with standard deviation $0.5$, but different
  mean values, namely, $(-1, 0)$, $(1, 0)$, $(0, -1)$ and $(0, 1)$;
  each mean is used to generate 25\% of the points.
\end{description}

 We use the Gaussian distribution to model a society with one dominant
 idea (e.g., where being moderate is the most popular position, or
 where a single dominant party exists). Since the boundary plays a significant role
 in the case of uniform distributions (we will discuss this
 effect below), we have chosen the Gaussian distribution, as its density
 vanishes close to the boundary.

 The 4-Gaussian distribution models a structured
 society, with four well-established positions (for the movie
 selection scenario, these might correspond to, e.g.,
 a combination of two genres and two typical budget values; in the
 world of politics, these could be four political parties).

 We also use the uniform distributions, on a square and on a disc, as
 intermediate cases, and in order to study specific behavior of voting
 rules at the border and, in case of the square, at the corners of the
 support of the distribution.

\begin{figure}[t!]
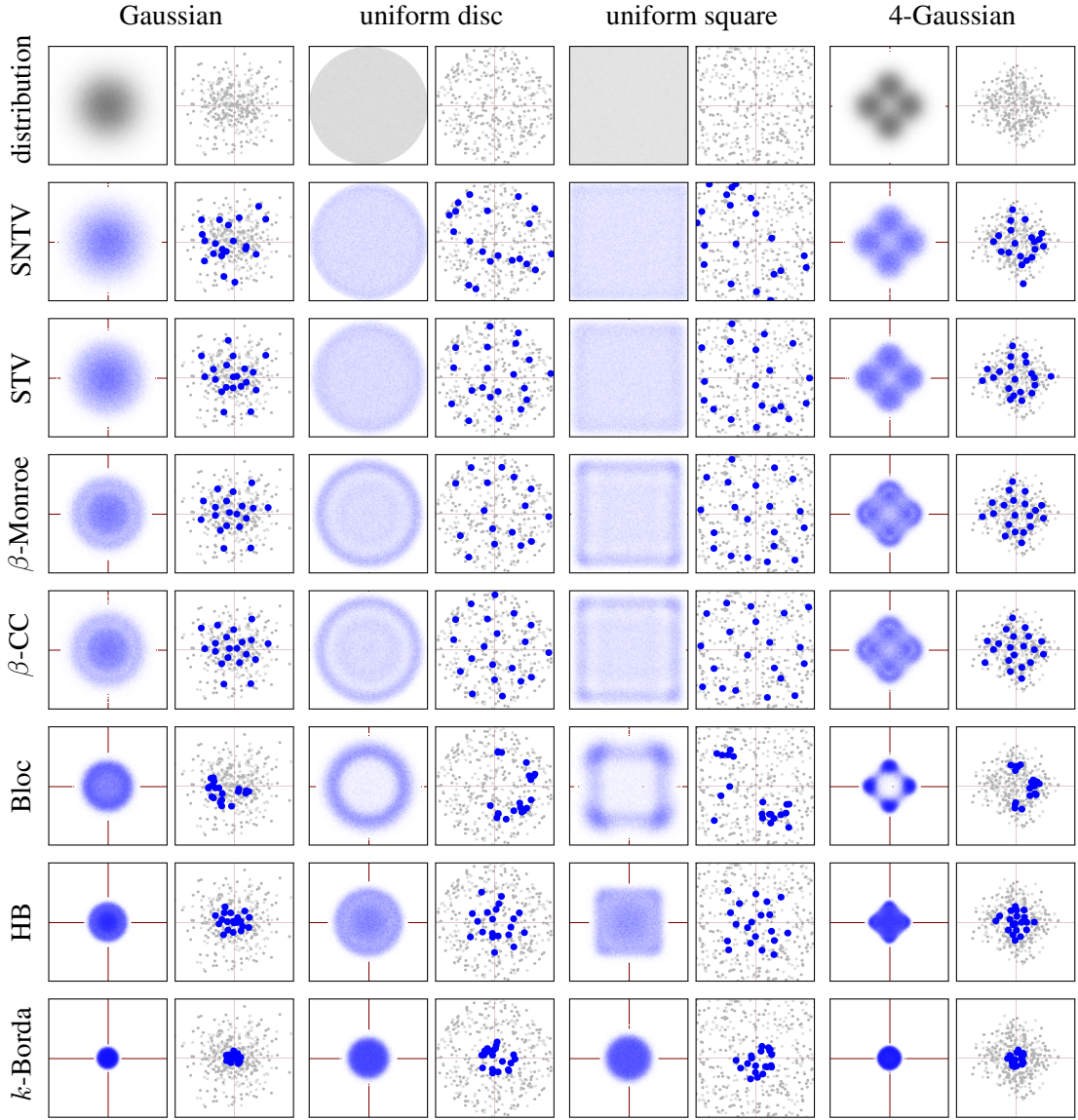

\noindent

\setlength{\tabcolsep}{1mm}
\centering
\begin{tabular}{ccccc}
 &  Gaussian & uniform disc & uniform square & 4-Gaussian \\
\resultsrule{distribution}{cdistrib}
\resultsrule{\;\;\;SNTV}{sntv}
\resultsrule{\;\;\;\;STV}{stv}
\resultsrule{$\beta$-Monroe}{monroeILP}
\resultsrule{\;\;\;\;$\beta$-CC}{ccILP}
\resultsrule{\;\;\;\;Bloc}{bloc}
\resultsrule{\;\;\;\;\;HB}{PAVtopk}
\resultsrule{\;\;$k$-Borda}{kborda}
\end{tabular}

\caption{\label{fig:main}Histograms and sample elections for our rules
  and distributions. The first row shows the distributions only.
  For sample election, voters are depicted as dark gray dots, candidates as light gray
  dots, and the winners as larger blue dots.}
\end{figure}

\pfparagraph{Raw Results.}   For each rule and each distribution, we have computed a
histogram, showing how frequently winners from a given location were
selected. These histograms, together with examples of elections and
their winning committees, are presented in Figure~\ref{fig:main} (the first
row presents the distributions themselves).

\begin{figure}[b!]
\centering
\includegraphics[width=8.cm]{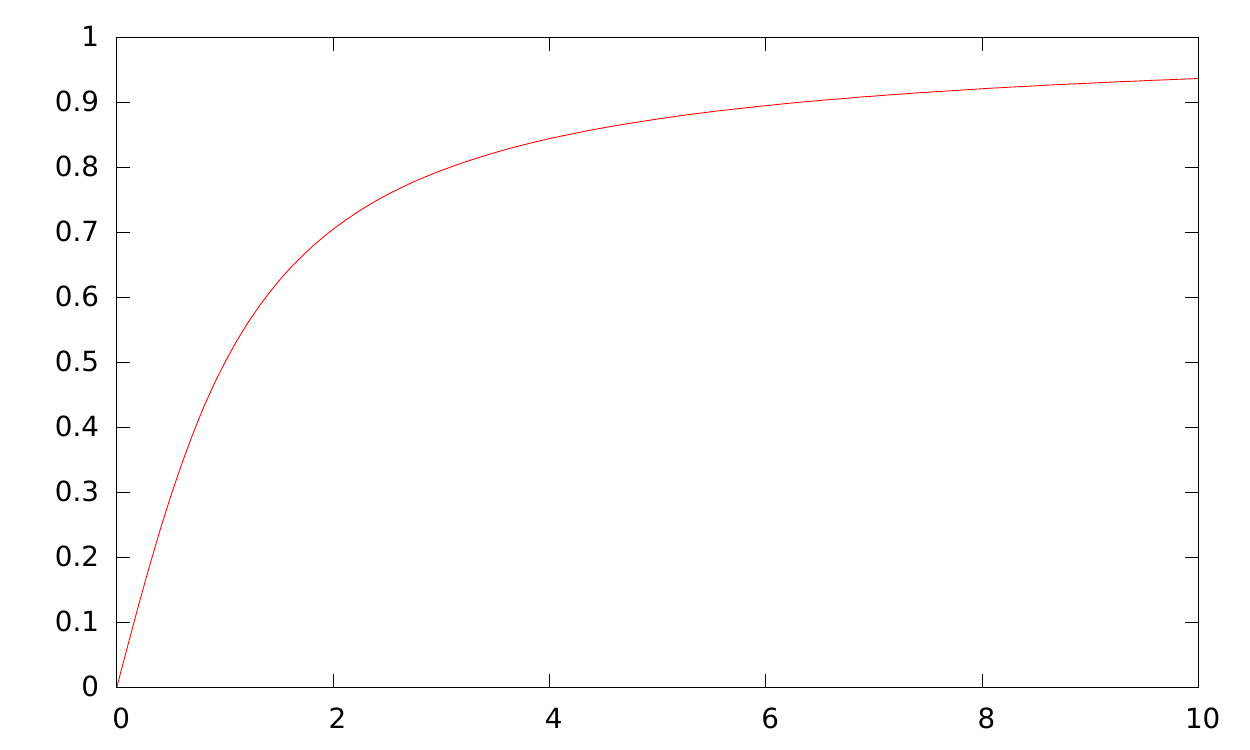}
\caption{\label{fig:arctan}Plot of the function $y =
  \frac{1}{\pi/2}\arctan\left( \frac{x}{\varepsilon T} \right)$ that
  we use for converting cell frequencies to color intensities.}
\end{figure}

The histograms were generated as follows. For each rule and
distribution, all the winners were always within the $[-3,3] \times
[-3,3]$ square. We have partitioned this square into $120 \times 120$
cells (each cell is a $0.05\times 0.05$ square), and---for each given
distribution and rule---counted how many times a member of the winning
committee fell into a given cell (we refer to this value as the
{\em frequency} of this cell). Then we have transformed the frequencies into
color intensities (the more winners fall into a particular cell, the
darker it is in Figure~\ref{fig:main}). Since there are big
differences among frequencies of cells across various rules and
distributions (e.g., the highest frequency of a cell for
$k$-Borda with the Gaussian 
distribution 
is over 27 times larger than the highest frequency of a cell for SNTV
under the uniform square distribution), we took the following
approach. Given a cell of frequency $x$, we compute its color
intensity $y$ ($0 \leq y \leq 1$; the closer is $y$ to $1$ the darker
is the cell) using the following formula:
\begin{equation}\label{eq:trans}
\ts  y = \frac{1}{\pi/2}\arctan\left( \frac{x}{\varepsilon T} \right),
\end{equation}
where $T$ is the sum of the frequencies of all the cells (so in our
case $T = 20 \cdot 10000$) and $\varepsilon$ is a parameter. We used
$\varepsilon = 0.0004$, so for the highest frequency of a cell in all
our experiments (found for $k$-Borda with the Gaussian distribution)
we have $x/(\varepsilon T) = 10.9$; for most other rules and
distributions this value is below $1.5$ and thus falls into the part
where our function behaves fairly linearly (see
Figure~\ref{fig:arctan}). To present the distributions themselves, we
computed histograms of the ideal points generated using our
distributions (on the technical side, to generate these histograms, we
used candidate positions from $10\,000$ generated elections for each
distribution; since formula~\eqref{eq:trans} is normalized, the
pictures in the first row of Figure~\ref{fig:main} are comparable to
those in the other rows).

\pfparagraph{Analysis.} We now consider the three applications of
multiwinner rules that we mentioned in the introduction and analyze
which of our rules are most suitable for each  application.

\pfparagraph{Parliamentary Elections.} We start with the case of
parliamentary elections. Intuitively, in this application we 
value proportional representation, which requires
that the distribution of the winners (as seen through the histograms) 
should be as close as possible to the underlying distribution of the voters.
Thus, at first sight, among our rules SNTV would be the
champion in this category. In addition, SNTV satisfies a number of
axioms studied by Elkind et
al.~\shortcite{elk-fal-sko-sli:j:multiwinner-properties}, especially
those geared towards proportional representation. However, at the same
time, it is intuitively clear that SNTV is not a very good rule
because it only takes the voters' top choices into
account, thus ignoring most of the information in the voters' preferences.
A look at the sample elections for SNTV (Figure~\ref{fig:main}) shows
that this intuition is correct: The reason why SNTV has such an
appealing histogram is that it selects committee members in areas
that, by random chance, have above-average density of voters 
and below-average density of candidates. Over all
$10\,000$ elections such areas are distributed evenly, similarly to
the distribution of the candidates and voters.

\begin{table}[t]
\begin{center}
\small
\begin{tabular}{l|cccc}
\toprule
rule & square & disc & Gauss. & 4 Gauss. \\
\midrule
{\bf SNTV}           & $\boldsymbol{3.292}$  & $\boldsymbol{3.219}$  & $\boldsymbol{3.275}$  & $\boldsymbol{2.787}$\\
STV            & $0.994$  & $1.070$  & $1.150$  & $1.043$\\
$\beta$-Monroe & $0.738$  & $0.797$  & $0.864$  & $0.765$\\          
$\beta$-CC     & $0.765$  & $0.820$  & $0.866$  & $0.826$\\
{\bf Bloc}           & $\boldsymbol{17.789}$ & $\boldsymbol{17.146}$ & $\boldsymbol{18.709}$ & $\boldsymbol{9.663}$\\
HB & $1.323$  & $1.391$  & $1.463$  & $1.289$\\
{\bf $\boldsymbol k$-Borda}      & $\boldsymbol{4.605}$  & $\boldsymbol{4.653}$  & $\boldsymbol{4.736}$  & $\boldsymbol{3.653}$\\
\bottomrule
\end{tabular}
\end{center}
\caption{\label{tab:quadrant}Variance of the number of winners in each quadrant. Bold font indicates
  rules where this value suggests asymmetric placement of winners on the plane (for $k$-Borda, this
  turns out to be a false alarm).}
\end{table}

This means that, in addition to considering the histograms, we also
need to check if results of individual elections are close to what the
histograms show. To this end we have used an indirect approach that,
nonetheless, turned out to be very effective. Let us fix some rule
$\calR$ and one of our distributions. For each generated election, we
(1) count how many members of the winning committee are in each of the
four quadrants $[0,\pm\infty)\times [0,\pm\infty)$, %
(2) collect these  numbers in a sequence, and (3) compute the variance of this sequence;
Table~\ref{tab:quadrant} shows the result of this computation, averaged over all instances. 
Since all our distributions are
symmetric with respect to the $x$ and $y$ axes, for rules that represent
voters proportionally \emph{in individual instances} we expect this number to be small.
Of course, the converse claim need not be true:
Low variance does not guarantee proportional representation.
That is, the variance-based approach can be used to eliminate `bad' rules 
rather than to identify `good' rules.

Table~\ref{tab:quadrant} 
clearly identifies a group of rules 
for which the variance of the number of winners per
quadrant is close to or below $1.0$, whereas for other rules the variance
is significantly higher (in our experiments, typically close to or above
$3.0$). Thus, the performance of SNTV (close to $3.0$) is a strong argument
against it.  On the other hand, the results for STV (both the shape of histograms
and the variance) indicate that it is an exceedingly good rule for selecting
parliaments. Indeed, this is the only rule with low variance that
is computationally tractable.
This is quite important, as STV is among
just a few nontrivial voting rules used in practice, 
yet some researchers---including some of us, until
recently---consider it unappealing.
The axiomatic results of Elkind et
al.~\shortcite{elk-fal-sko-sli:j:multiwinner-properties} and our
experiments 
provide different arguments in favor of using STV for proportional representation.

The results for $\beta$-Monroe are slightly less appealing than those
for STV. While the variance of the number of winners per quadrant is
low, the histograms are farther from resembling the distributions of
candidates and voters. They are very similar to those for $\beta$-CC, which
should not be too surprising.
In our experiments, the only difference between these rules is that
$\beta$-Monroe is forced to assign exactly $10$ voters to each
selected committee member, whereas $\beta$-CC can choose an optimal assignment,
where the number of voters assigned to each committee member may be arbitrary.
Nonetheless, for each of the distributions, around 80\% of
the committee members selected by $\beta$-CC were assigned to between
$7$ and $13$ voters each. In effect, the assignments computed by
$\beta$-CC and $\beta$-Monroe were quite similar. 
Naturally, if the distributions of candidates and voters were not
identical, the results would be different as well (we have run initial
experiments to confirm this, available in the appendix). Below we discuss the intriguing patterns in the histograms for
$\beta$-CC (a similar explanation applies to $\beta$-Monroe).

\pfparagraph{Portfolio/Movie Selection.}Let us now consider the
portfolio/movie selection
scenario~\cite{bou-lu:c:chamberlin-courant,bou-lu:c:value-directed-cc,elk-fal-sko-sli:j:multiwinner-properties,sko-fal-lan:j:collective}. Here
we care mostly about the diversity of the committee and, intuitively,
we would like to obtain histograms that cover a large chunk of the
support of the distribution, but which---as compared to the parliamentary elections
setting---are less responsive to the densities of the candidates and
voters.

We first analyze the results for $\beta$-CC, a rule that seems to be
designed exactly for this scenario. However, it does not quite
fit the description above. As we will see, to some extent this is due to
the nature of the rule, and to some extent this is because our initial
expectations were not entirely reasonable. There are two main issues
regarding $\beta$-CC.

The first one concerns what we call the \emph{edge effect} and the
\emph{corner effect}.  Let us consider the uniform square
distribution. If a candidate is located 
far from the edges, then he or she is also surrounded by a relatively
large number of other candidates with whom he or she needs to compete
for a high position in the voters' preference orders. On the other hand,
if a candidate is located near an edge (or, better yet, near a corner)
then the competition is less stiff.
However, if a candidate is close to the edge/corner,
the number of voters for whom he or she would be a representative
also decreases. In effect, for the uniform square and uniform disc
distributions, we see increased frequencies of winners near (but not
exactly on) the edges and corners.
The edge and corner effects are
visible also for SNTV and STV (though to a lesser extent), and 
they are very prominent for Bloc (especially in conjunction with cases where
an area near edge/corner has an above-average density of
voters). %

The second issue regarding $\beta$-CC is that when some candidate is
included in the committee, other candidates that are very close to him or
her are unlikely to be selected; indeed, this behavior is
quite desirable when one wants to maintain diversity of the
committee.  This explains why for the uniform square and uniform disc
distributions the near-edge area with increased frequencies is
surrounded by an area with lower frequencies. This effect also
explains the interesting pattern for the 4-Gaussian
distribution. Since there are many voters in the centers of the four
Gaussians, candidates from these locations are likely to be included
in the committee. But this very fact strongly decreases the chances of
the candidates that are located just a bit further away
from the centers of the Gaussians.

Our visual inspection of the election results for
$\beta$-CC shows that every single committee appears to be diverse and
appealing for the portfolio/movie selection problem (this is also
supported by the low value of the variance of the number of winners
per quadrant). However, the histograms show that the rule also has an
implicit, systematic bias against certain candidates (the nature of
this bias depends on the distribution) that users of the rule 
should
take into account.

HarmonicBorda also appears to be a very interesting rule for the
portfolio/movie selection task (and, perhaps, even for parliamentary
elections). In our experiments, HarmonicBorda chose committees
distributed fairly uniformly in the central areas, ignoring candidates
with extreme opinions.

\pfparagraph{Shortlisting.} %
Here our guiding principle is that the committee should consist of
similar candidates (i.e., located close to each other). For this
criterion, $k$-Borda is our rule of choice. In all of the experiments
it consistently chose candidates located in the center, close to each
other. %
Table~\ref{tab:quadrant} indicates that $k$-Borda has high variance of the number
of winners per quadrant. We believe that this is caused not by any faults
of the rule itself,
but by a fairly natural statistical property of our distributions.
Since $k$-Borda selects $20$ candidates from the center, due to random
perturbations, sometimes the central candidates are not distributed
over the quadrants in a perfectly balanced way, and 
our variance-based measure does not take into account
the candidates' centrality.

\pfparagraph{The Strange Case of Bloc.} In the
situation where $k$ candidates are to be selected (e.g., to a city
council), it is quite common to ask the voters to come up with $k$ names (ranked or
non-ranked). Bloc, in particular, is quite a popular rule. Our
histograms show that Bloc is very sensitive to the edge and corner
effects (the pattern is similar to that for $\beta$-CC, but the effects are much stronger). 
Worse yet, Table~\ref{tab:quadrant} shows very high variance of the number
of winners in each quarter and, indeed, the example elections for Bloc
in Figure~\ref{fig:main} show very asymmetric placements of the
winners. These two arguments by themselves make Bloc a questionable voting
rule.

Bloc is also the only rule in our collection that shows the following
\emph{inversion} effect: For the Gaussian distribution, the
frequencies of the cells near the center (i.e., near the mean of the
Gaussian distribution) are lower than the frequencies of the cells in
the ring surrounding it. This is a very counter-intuitive and unexpected
phenomenon: The most popular views in the society are represented less
frequently than the not-so-popular ones. We believe that the
mechanism behind this effect is similar to that behind the edge/corner
effect: Even though the center has the highest density of the voters,
it also has the highest density of the candidates, who therefore
``steal points away'' from each other. As a consequence, the slightly less
popular candidates in the ring get enough support (both
from some of the voters in the center and from those on the ring and
beyond) to be elected.\footnote{Indeed, this can be seen as a type of approximate cloning
(see the discussion in the papers of of Tideman~\shortcite{tid:j:clones}, Laffond et
al.~\shortcite{laf-lai-las:j:composition}, and Elkind et
al.~\shortcite{elk-fal-sli:j:cloning}).}

\newcommand{\sizeshistogram}[2]{\fbox{\includegraphics[width=1.65cm]{hist_bloc/#1_#2_hist.png}}}

\newcommand{\resultsrulesizes}[2]{
\rotatebox{90}{#1} &
\sizeshistogram{#2}{10} &
\sizeshistogram{#2}{20} &
\sizeshistogram{#2}{30}\\[0mm]
}

\begin{figure}[t!]
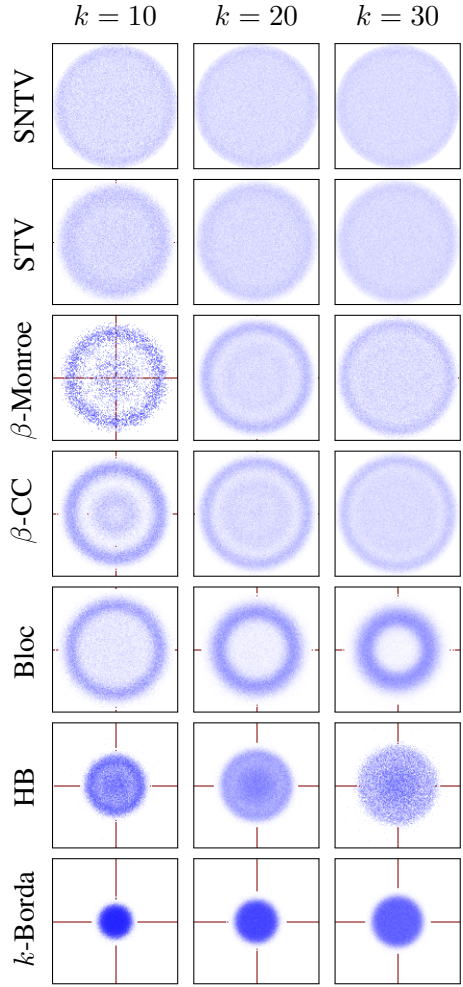

  \centering
  \setlength{\tabcolsep}{1mm}
  \begin{tabular}{cccc}
   & $k=10$ & $k=20$ & $k=30$ \\[1mm]
\resultsrulesizes{\;\;\;SNTV}{sntv}
\resultsrulesizes{\;\;\;\;STV}{stv}
\resultsrulesizes{$\beta$-Monroe}{monroeILP}
\resultsrulesizes{\;\;\;\;$\beta$-CC}{ccILP}
\resultsrulesizes{\;\;\;\;Bloc}{bloc}
\resultsrulesizes{\;\;\;\;\;HB}{PAVtopk}
\resultsrulesizes{\;\;$k$-Borda}{kborda}
  \end{tabular}

 \caption{\label{fig:sizes}Histograms for our rules under the disc
     distribution, for committee sizes $10$, $20$, and $30$. For
     HarmonicBorda ($k \in \{10,30\}$) and Monroe ($k=30$) we computed
     only 5000 elections. Due to technical issues, for $\beta$-Monroe
     with $k=10$ we computed only about 500 elections.}

\end{figure}

\section{Robustness of the Results}

So far we have considered elections with $m = 200$ candidates, $m =
200$ voters, and committee size $k=20$ only. Thus it is natural to
wonder if our conclusions remain valid as we vary these parameters.

Except for STV and $\beta$-Monroe, all our rules belong to the class
of committee scoring
rules~\cite{elk-fal-sko-sli:j:multiwinner-properties,fal-sko-sli-tal:j:committee-scoring-rules-hierarchy},
i.e., they define a per-voter score of each possible committee and
select committees for which the sums of these scores are the
highest. In consequence, the results for these rules should not change
significantly with the number of voters (unless this number becomes
very small). Since STV and $\beta$-Monroe are similar in spirit to
committee scoring rules (indeed, STV is similar to SNTV and
$\beta$-Monroe is very closely related to $\beta$-CC), the
results %
for them should be similarly robust.

We also do not expect strong qualitative differences in our results
for different numbers of candidates or different committee sizes
(again, except for very small values). Nonetheless, we do observe
quantitative differences.  

In Figure~\ref{fig:sizes} we present
histograms for our rules with respect to the disc distribution, for committee
sizes $10$, $20$, and $30$
(the histogram
for committee size $20$ is the same as in Figure~\ref{fig:main}; we
repeat it for the sake of comparison). We note that the results for
SNTV and STV are nearly the same irrespective of the committee
size.\footnote{
For $k=30$, the quota for STV is
  $q = \lfloor \frac{200}{31}\rfloor +1 = 7$. Thus, in the first
  28 stages we remove 196 voters, 
  so the 29th candidate is chosen by 4 voters and the 30th candidate is selected randomly.}   
The results for Bloc, HarmonicBorda, and $k$-Borda also look
very similar, and the differences are only in the radii of the
discs/rings generated by these rules (this is especially natural for
$k$-Borda; as we choose more and more of the centrally located
candidates, they form a larger and larger disc).  The results for
$\beta$-CC and $\beta$-Monroe for different committee sizes also look
similar, but for $k=10$ (especially for the case of $\beta$-CC) the
artifacts in the histograms become much more visible (e.g., for $k=10$
and $\beta$-CC, there are two very clearly visible consecutive rings).
This indicates that our observations about $\beta$-CC and
$\beta$-Monroe do not necessarily carry over to the case of very small
committees.

\section{Conclusions}

Our results lead to several interesting observations. Foremost, 
within the framework of our study STV stands out as an exceptionally
good rule for parliamentary elections. On the other hand, the Monroe
rule, which is also an appealing rule for this application, did not do
quite as well. We also found that the Monroe and Chamberlin--Courant
rules may have (somewhat surprising) implicit biases against some
candidates. Further, we %
discovered that in our experiments HarmonicBorda tends to ignore
extremist candidates and fairly uniformly covers central areas (this
seems quite related to the results of Aziz et
al.~\shortcite{azi-bri-con-elk-fre-wal:j:justified-representation} on
justified representation).  We confirmed that $k$-Borda has good
properties as a shortlisting rule and
provided %
strong arguments against the Bloc rule.

\pfparagraph{Acknowledgments.}  Edith Elkind and Piotr Skowron were
supported by the ERC grant 639945 (ACCORD), Piotr Faliszewski was
supported by the NCN grant 2016/21/B/ST6/01509, Arkadii Slinko was
supported in part by the Marsden Fund 3706352 of The Royal Society of
New Zealand, and Nimrod Talmon was supported by a postdoctoral
fellowship from I-CORE ALGO. Jean-Fran{\c c}ois Laslier thanks the ANR
project ANR13-BSH1-0010 DynaMITE.

\bibliography{grypiotr2006}

\appendix

\section{Overview}

In this appendix we present %
the results %
omitted from the main part of the paper.
First, we present approximation algorithms for the Chamberlin--Courant
and Monroe rules (including one that is due to this paper) and discuss
the results for them. Then we show our preliminary results for a
scenario where the distributions of candidates and voters are not
similar.  Finally, we show our Integer Linear Program (ILP)
formulation for HarmonicBorda.

\begin{figure}[t!]
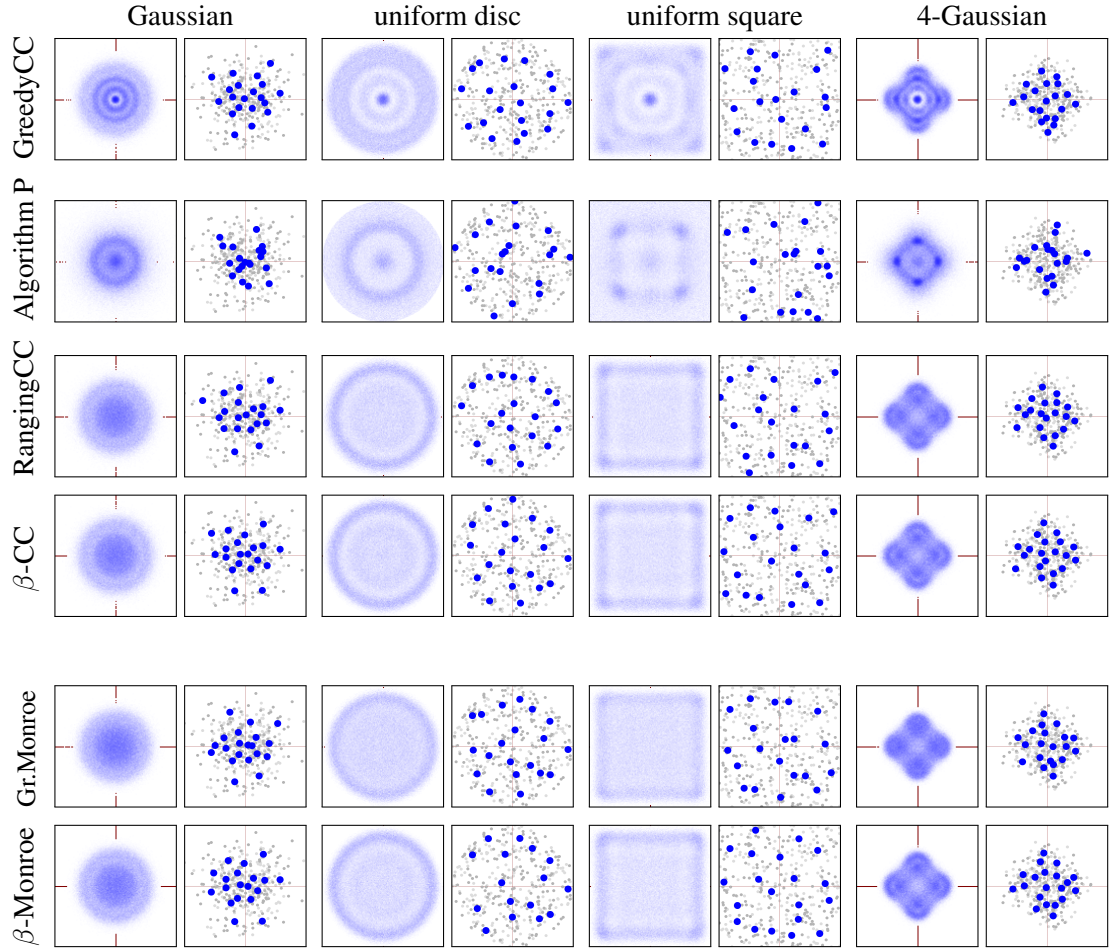

\noindent

\setlength{\tabcolsep}{1mm}
\begin{tabular}{ccccc}
 & Gaussian & uniform disc  & uniform square & 4-Gaussian \\

\resultsrule{GreedyCC}{greedyCC}
\resultsrule{Algorithm~P}{algoP}
\resultsrule{RangingCC}{algoP_ranging}
\resultsrule{\;\;\;$\beta$-CC}{ccILP}
\\[2mm]
\resultsrule{\small Gr.Monroe}{greedyMonroe}
\resultsrule{$\beta$-Monroe}{monroeILP}
\end{tabular}
\caption{\label{fig:approx}Results for approximation algorithms for
  $\beta$-CC and $\beta$-Monroe.}
\end{figure}

\section{Approximation Algorithms}

Let us first consider approximation algorithms for $\beta$-CC. Recall
that if $E = (C,V)$ is an election, $k$ is a committee size, and
$\Phi$ is a $k$-CC-assignment, then by $\beta(\Phi)$ we denote the sum
of Borda scores that the voters assign to their representatives
(with respect to $\Phi$). Given a committee $W$ and election $E$, by
the \emph{grab-your-best} assignment of $W$ to the voters in $E$ we
mean the function $\mathrm{gyb}(W,E)$ which assigns to each voter the member of $W$
which this voter ranks highest.

We consider the following three approximation algorithms for
$\beta$-CC (we use the same notation as in the description of our
multiwinner rules; $E = (C,V)$ is the election at hand and $k$ is the
committee size):

\begin{description}
\item[GreedyCC.] The algorithm starts by setting the initial committee
  $W$ to be empty, and then executes the following $k$ iterations: In
  each iteration, it extends the committee $W$ with a candidate $c$
  (previously not included in $W$) that maximizes
  $\beta(\mathrm{gyb}(W \cup \{c\}))$. (In particular, the algorithm
  always starts by including the candidate with the highest Borda
  score.) Finally, it outputs the computed committee $W$. GreedyCC is
  due to Lu and Boutilier~\shortcite{bou-lu:c:chamberlin-courant} and
  guarantees approximation ratio of at least $1-\frac{1}{e} \approx
  0.63$.

\item[Algorithm~P.] This algorithm proceeds as follows. First, it
  computes a threshold value $x = \frac{|C| \mathrm{w}(k)}{k}$ (where
  $\mathrm{w}(\cdot)$ is Lambert's $\mathrm{w}$ function; $\mathrm{w}(k)$ is
  $o(\log k)$). Then it sets the initial committee $W$ to be empty and
  executes $k$ iterations as follows: In each iteration, it finds a
  candidate $c$ that is ranked among the top $x$ positions by the largest
  number of voters.  Then it adds $c$ to $W$ and deletes all the
  voters that rank $c$ among their top $x$ positions. (Thus, the algorithm
  can be seen as an incarnation of a greedy SetCover algorithm, where
  voters are items to be covered and each candidate covers those
  voters that rank him or her among their top $x$ positions). Finally, it
  outputs $W$. The algorithm is due to Skowron et
  al.~\shortcite{sko-fal-sli:j:multiwinner} and achieves approximation
  ratio of $1 - \frac{2\mathrm{w}(k)}{k}$.
  We mention that it is also a basis of a polynomial-time approximation scheme for $\beta$-CC.

\item[RangingCC.] This is an extension of Algorithm~P introduced in this
  paper. RangingCC computes the committees using Algorithm~P for
  threshold values $x$ between $1$ and $\frac{|C| \mathrm{w}(k)}{k}$ and
  outputs the one with the highest $\beta$-CC score.
\end{description}

For $\beta$-Monroe, we consider the GreedyMonroe algorithm of Skowron
et al.~\shortcite{sko-fal-sli:j:multiwinner};\footnote{In the paper of
  Skowron et al., it is denoted as Algorithm~A.}  again, we use the
same notation as in the description of multiwinner rules above (so we
seek $k$ winners for election $E = (C,V)$ with $m$ candidates and $n$
voters):
\begin{description}
\item[GreedyMonroe.] The algorithm starts by setting $W$ to be the
  empty committee. Then it constructs a Monroe assignment iteratively
  as follows (for simplicity, let us assume that $k$ divides
  $n$). At the beginning of each iteration, the algorithm finds a
  candidate $c$ and $\frac{n}{k}$ voters, denoted by $V(c)$, that jointly maximize
  the Borda score of $c$ in the election $(C,V(c))$. Then, the algorithm adds
  $c$ to $W$, assigns $c$ to each voter in $V(c)$, and removes the
  voters in $V(c)$ from further considerations.  The algorithm guarantees
  an approximation ratio of
  $1-\frac{k-1}{2(m-1)}-\frac{H_k}{k}$, where $H_k =
  1+\frac{1}{2}+\cdots +\frac{1}{k}$ is the $k$'th harmonic number.
\end{description}

\begin{table}[b]
\begin{center}
\small
\begin{tabular}{l|cccc}
\toprule
rule & square & disc & Gauss. & 4 Gauss. \\
\midrule
GreedyCC          & $1.019$  & $1.083$  & $1.106$  & $1.132$\\
Algorithm~P         & $2.551$  & $2.453$  & $2.418$  & $2.381$\\
RangingCC      & $0.907$  & $0.944$  & $1.015$  & $0.959$\\
GreedyMonroe      & $0.848$  & $0.926$  & $0.978$  & $0.877$\\
\bottomrule
\end{tabular}
\end{center}
\caption{\label{tab:quadrant-appendix}Variance of the number of winners in each quadrant.}
\end{table}

\section{Results for Approximation Algorithms}

The histograms for the approximation algorithms are presented in
Figure~\ref{fig:approx} (together with the repeated histograms for
$\beta$-CC and $\beta$-Monroe) and their variances for the number of
winners per quadrant are in Table~\ref{tab:quadrant-appendix}.

\pfparagraph{Approximation Algorithms for $\boldsymbol \beta$-CC.} The
results of the approximation algorithms for $\beta$-CC are rather
varied, but even a quick glance shows that RangingCC seems to be the
closest to the original $\beta$-CC rule. While we provide explanation
as to why the other two algorithms are not doing well, the performance of
RangingCC came as a surprise to us and we still do not have a very good explanation for its behavior.

To understand the behavior of GreedyCC, it suffices to recall
that---by definition---in the first iteration the algorithm chooses
the Borda winner. In our elections, the Borda winner is always located
very close to the center, so the histograms for GreedyCC show a spike
there. Then, due to the nature of the Chamberlin--Courant rule (as
described in the main body of the paper), the algorithm selects
candidates that are not too close to this first winner. This explains
the patterns that we see for all our distributions. These patters are
far more visible for GreedyCC than for $\beta$-CC, in particular,
because the first iteration chooses a candidate from almost the same
location irrespective of the actual distribution of the points.
GreedyCC achieves good results for the variance for the number of
winners per quadrant.

The behavior of Algorithm~P can be explained similarly to that of
GreedyCC, but by an analogy to the Bloc rule. Algorithm~P considers
candidates ranked at the top $x$ positions, where $x$ is a prespecified
threshold value (recall the description of the algorithm). In effect,
the first iteration is almost the same as in Bloc, except that Bloc
chooses a candidate ranked most frequently among the top $k$ positions and
Algorithm~P considers the top $x$ positions. In the second iteration,
Algorithm~P chooses a candidate that is ranked among the top-$x$ positions
by many voters who are far from the candidate chosen in the first
iteration. Such a candidate is likely to also be included in the Bloc
committee (again, taking into account the fact that both rules
consider slightly different numbers of top candidates). We believe
that similar effect lasts for a few iterations and is sufficient to
create those patterns in the histograms of Algorithm~P which resemble
Bloc. However, in further iterations Algorithm~P starts behaving
differently than Bloc and, for example, chooses candidates from the
center (especially for the Gaussian and uniform disc distributions).
Unfortunately, Algorithm~P has poor variance of the number of winners
per quadrant (on the order of $2.4$-$2.5$) and, indeed, visual
inspection of its results shows that they are not satisfactory. Thus
we believe that it should not be used (even though, in most settings, its guaranteed
approximation ratio is better than that of GreedyCC).

Finally, RangingCC achieves nearly the same histograms as $\beta$-CC
and has very good results for the variances for the number of winners
per quadrant (but still slightly higher than $\beta$-CC). Since
RangingCC winners can be computed quite efficiently, it appears to be
the best choice among the three algorithms we have tested (in
practice, one might also try the clustering technique of~\cite{faliszewski2016achieving}. Nonetheless, we are quite baffled with the performance of
RangingCC and do not really have convincing explanations for its superiority against its component algorithms
(various incarnations of Algorithm~P).

\newcommand{\overlappingsquares}[2]{\rotatebox{90}{#1} &
\histogram{overlapping_squares}{#2}
\samplerun{overlapping}{#2}\\
}

\begin{figure}[t!]
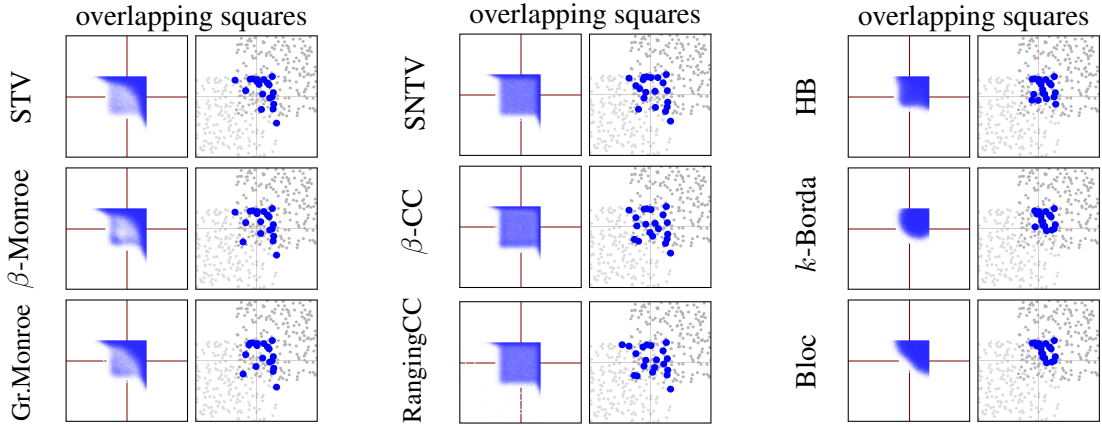

\centering
\begin{tabular}{cc}
& overlapping squares \\
  \overlappingsquares{\;\;\;\;STV}{stv}
  \overlappingsquares{$\beta$-Monroe}{monroeILP}
  \overlappingsquares{\small Gr.Monroe}{greedyMonroe}
\end{tabular}\hspace{6mm}
\begin{tabular}{cc}
& overlapping squares \\
  \overlappingsquares{\;\;\;SNTV}{sntv}
  \overlappingsquares{\;\;\;\;$\beta$-CC}{ccILP}
  \overlappingsquares{\small RangingCC}{algoP_ranging}
\end{tabular}\hspace{6mm}
\begin{tabular}{cc}
& overlapping squares \\
  \overlappingsquares{\;\;\;\;\;HB}{PAVtopk}
  \overlappingsquares{\;\;$k$-Borda}{kborda}
  \overlappingsquares{\;\;\;\;Bloc}{bloc}
\end{tabular}

\caption{\label{fig:overlapping}Histograms and sample elections for several voting rules, for
  the case where the candidates and voters are distributed uniformly
  on two overlapping squares.}
\end{figure}

\pfparagraph{GreedyMonroe.}  It appears that GreedyMonroe is a very
good approximation algorithm for $\beta$-Monroe. The histograms we
obtained for it are very similar to those for $\beta$-Monroe and the
variance for the number of winners per quadrant is low (if a bit higher
than for $\beta$-Monroe).  In fact, GreedyMonroe's histograms appear
to be a bit more similar to the underlying distributions of candidates
and voters than those of $\beta$-Monroe, which by our criteria makes
it a slightly better rule for parliamentary elections than the latter.
Indeed, this also shows in the results of Elkind et
al.~\shortcite{elk-fal-sko-sli:j:multiwinner-properties}, who prove
that GreedyMonroe satisfies the solid coalitions property---a property
desirable for proportional representation\footnote{Formally, the
  property says the following: if we have an election with $n$ voters, we want to
  choose a committee of size $k$, and there is a candidate $c$ that is
  ranked on the first place by at least $\frac{n}{k}$ voters, then
  this candidate shall be included in the winning committee.}---and that
$\beta$-Monroe does not. Interestingly, while Elkind at
al.~\shortcite{elk-fal-sko-sli:j:multiwinner-properties} did not
insist strongly on this property, our three rules with histograms most
similar to the underlying distributions (STV, SNTV, and GreedyMonroe)
do satisfy it.

\section{Overlapping Squares Distribution}

So far, our results for $\beta$-Monroe and $\beta$-CC were quite
similar.  To show that the rules are, indeed, different, we have
performed a quick experiment for a setting where the distributions of
the ideal points of candidates and voters are not the same:
\begin{description}
\item[Overlapping Squares.] The ideal points of the candidates are
  distributed uniformly on the $[-3,1] \times [-3,1]$ square, whereas
  the ideal points of the voters are distributed uniformly on the
  $[-1,3]\times[-1,3]$ square.
\end{description}
Naturally, we should not expect any society to really follow such a
distribution and we use it only as a test case.

It turns out that STV, SNTV, $\beta$-CC, $\beta$-Monroe, RangingCC,
and GreedyMonroe can be partitioned into two groups. STV,
$\beta$-Monroe and GreedyMonroe aim for proportional representation
and, thus, their histograms put more emphasis on the candidates near
the $(1,1)$-corner. $\beta$-Monroe also puts some emphasis on the
$(-1,-1)$ corner, while GreedyMonroe and STV do it only to a very
minor extent. On the other hand, SNTV, $\beta$-CC, and RangingCC are
more geared towards covering the intersection of the supports of the
distributions of candidates and voters. We view this as further evidence that
these rules (or, rather, only $\beta$-CC and RangingCC, since we
already argued against SNTV) are well-suited for portfolio/movie
selection tasks.

As to the three other rules, HarmonicBorda, $k$-Borda, and Bloc, note
that they concentrate on a support that is strictly smaller than the
intersection of the two distributions, and tilted towards the center
of the voters' distribution. This confirms the tendency of these rules
to be detrimental to extreme candidates.

\section{Integer Linear Program for HarmonicBorda}

In this section we describe the integer linear program that we have
used for computing HarmonicBorda.

Let $E = (C,V)$ be an input election with $m$ candidates and $n$
voters.  We are interested in a winning committee $S$ of size $k$.  We
define the following binary variables.  For $j \in [m]$, we define
$x_j$, with the intent that $x_j = 1$ if and only if $c_j \in S$.  For
$i \in [n], j \in [m], \ell \in [k]$, we define $y_{i, j}^\ell$, with the
intent that $y_{i, j}^\ell = 1$ if and only if the $j$th-ranked candidate
of voter $v_i$ is chosen as her $l$-th best committee member.  We have
the following optimization goal:
\[
  \max {\sum_{i \in [n]} \sum_{j \in [m]} \sum_{\ell \in [k]}} \frac{1}{\ell} \cdot \beta_m(j) \cdot y_{i, j}^\ell;
\]
and we include the following constraints: 
\begin{enumerate}
\item The committee includes exactly $k$ candidates:
\[
  \sum_{j \in [m]} x_j = k.
\]

\item For a given voter and position $j$, the candidate on position
  $j$ can be $\ell$-th best committee member for this voter for at
  most one value of $\ell$. Formally, for each $i \in [n]$ and for
  each $j \in [m]$, we have the constraint:
\[
  \sum_{\ell \in [k]} y_{i, j}^\ell \leq 1;
\]

\item For a given voter, there is exactly one candidate that this
  voter ranks as $\ell$-th best in the committee. Formally, for each
  $i \in [n]$ and for each $\ell \in [k]$, we have the constraint:
\[
  \sum_{j \in [m]} y_{i, j}^\ell = 1;
\]

\item A candidate cannot be the $\ell$-th best committee member for a
  given voter if this candidate is not even a committee
  member. Formally, for each $i \in [n]$, for each $j \in [m]$, and
  for each $\ell \in [k]$, we have the constraint:
\[
  y_{i, j}^\ell \leq x_t;
\]
where $c_t$ is the $j$-th ranked candidate of voter $v_i$.
\end{enumerate}

\end{document}